\documentclass[twocolumn,english,aps,prl,floatfix,amssymb,superscriptaddress]{revtex4-1}
\usepackage[T1]{fontenc}
\usepackage[utf8]{luainputenc}
\usepackage{amsmath}
\usepackage{amssymb}
\usepackage{graphicx}
\usepackage{esint}

\makeatletter
\@ifundefined{textcolor}{}
{%
 \definecolor{BLACK}{gray}{0}
 \definecolor{WHITE}{gray}{1}
 \definecolor{RED}{rgb}{1,0,0}
 \definecolor{GREEN}{rgb}{0,1,0}
 \definecolor{BLUE}{rgb}{0,0,1}
 \definecolor{CYAN}{cmyk}{1,0,0,0}
 \definecolor{MAGENTA}{cmyk}{0,1,0,0}
 \definecolor{YELLOW}{cmyk}{0,0,1,0}
}

\usepackage{babel}
\usepackage[bookmarks=false,linkcolor=blue,urlcolor=blue,colorlinks,citecolor=blue]{hyperref}
\makeatother

\usepackage{babel}
\begin{document}

\title{Supersymmetry in the Fractional Quantum Hall Regime}
\author{Eran Sagi}
\affiliation{Department of Condensed Matter Physics, Weizmann Institute of Science, Rehovot 76100, Israel}
\author{Raul A. Santos}
\altaffiliation{Current address: School of Physics \& Astronomy, University of Birmingham, Edgbaston, Birmingham, B15 2TT, United Kingdom}
\affiliation{Department of Condensed Matter Physics, Weizmann Institute of Science, Rehovot 76100, Israel}

\begin{abstract}
Supersymmetry (SUSY) is a symmetry transforming bosons to fermions
and vice versa. Indications of its existence have been extensively sought after in high-energy experiments.
However, signatures of SUSY
have yet to be detected. In this manuscript we propose a condensed matter realization of SUSY on the edge of a Read-Rezayi
quantum Hall state, given by filling factors of the form $\nu =\frac{k}{k+2}$, where $k$ is an integer. As we show, this strongly
interacting state exhibits an $\mathcal{N}=2$ SUSY. This allows us to
use a topological invariant - the Witten index - defined specifically for supersymmetric
theories, to count the difference between the number of bosonic and fermionic zero-modes in a circular edge. In our system, we argue that the edge
hosts $k+1$ protected zero-modes. We further discuss the stability of SUSY with respect to generic perturbations, and find that much of the above results remain unchanged. In particular, these
results directly apply to the well-established $\nu=1/3$ Laughlin state, in which case SUSY is a highly robust property of the edge theory.
These results unveil a hidden topological structure on the long-studied Read-Rezayi states.
\end{abstract}
\maketitle

{\it Introduction:} Since its discovery, the quantum Hall effect has led to a plethora
of remarkable new physical phenomena. The integer quantum Hall (IQH) effect
 \cite{Klitzing1980}, for instance, is a paradigmatic example of non-interacting
topological phases, characterized by bulk topological invariants and
gapless edge modes. The strongly interacting fractional quantum Hall
(FQH) states \cite{Tsui1982}, on the other hand, present even more striking properties,
such as the existence of fractionally charged anyonic bulk excitations.

A subset of the fractional states are non-Abelian ones, whose bulk
excitations are non-Abelian anyons, and whose edges realize non-trivial
interacting conformal field theories (CFTs).
While the recent interest in non-Abelian phases is mostly driven by their
exotic bulk excitations and the possibility of using them as resources in topological quantum computation, these states are also a
natural playground for experimentally studying one-dimensional (1D) conformal field theories (CFTs).

In this manuscript we will study the edge CFTs of Read-Rezayi (RR) states at filling $\nu=\frac{k}{k+2}$. The simplest
example is given by the $\nu=1/3$ Laughlin state, corresponding to $k=1$, which constitute the most prominent FQH
state  \cite{Tsui1982}. RR states with $k>1$, on the other hand, are widely believed to be energetically unfavorable
compared to other competing states at the same filling factors in the lowest Landau-level. In
the first excited Landau-level, however, numerical works indicate that the particle-hole conjugates of these states may be the ground-states in the corresponding
filling factors \cite{Read1999,Storni2010,Zaletel2015,Zhu2015,Mong2015}. Indeed, the plateaus observed at $\nu=5/2$ and $\nu=12/5$ are
strong candidates for realizing the particle-hole conjugates of the $k=2$ and $k=3$ states.
As we will show, supersymmetry (SUSY) - a symmetry transforming bosons to fermions and vice versa - emerges naturally in these
states.

In general, SUSY is a space-time symmetry which constitutes the only possible extension of the Poincaré
group consistent with the symmetries of the scattering-matrix \cite{Haag1975}. It has attracted attention given
that it solves several open problems in high-energy physics and cosmology \cite{Dimopoulos1981,WITTEN1981,DINE1981,Dimopoulos1981b,Sakai1981,KAUL1982}.
In particular, its existence implies that the strengths of the three fundamental forces
of the standard model unify at the same energy scale \cite{baer2006}. Furthermore, if it is in fact a symmetry of nature, it will provide
natural candidates for dark matter particles.

Despite its many features, the existence of this symmetry has not
been confirmed in high energy experiments so far. This has recently sparked interest in realizing SUSY in condensed matter systems.
In particular, signatures of space-time SUSY have been proposed in the spontaneous time-reversal symmetry breaking transition
on the edge of topological systems \cite{Grover2012,Grover2014}. Specifically, it was shown that in this case the critical point
belongs to the same universality class as the tricritical Ising model, and therefore possesses $\mathcal{N}=1$ SUSY. The same
universality has been discovered in strongly interacting Majorana chains \cite{Rahmani2015}.
Recently \cite{Hsieh2016}, $\mathcal{N}=2$ SUSY was shown to generically exist in
translation invariant lattice systems with an odd number of Majorana degrees of freedom per unit cell.
It has also been suggested that $\mathcal{N}=2$ SUSY appears at the tricritical point between a Dirac semi-metal and a
pair-density-wave on the surface of a correlated topological insulator \cite{Jian2016}.

In this manuscript, we demonstrate that the low-energy description of the edge of RR states gives rise to an $\mathcal{N}=2$ SUSY,
generated by two fermionic charges $Q_1,Q_2$. Combining this with the existence of conformal symmetry,
we find that the edge of our incompressible state realizes an $\mathcal{N}=2$ superconformal theory in 1+1
spacetime dimensions.

Once we establish the emergence of SUSY in our quantum Hall system,
we will turn to study its implications. To do so, we will introduce the so-called Witten index, which
is a fundamental topological invariant measuring the number of bosonic
 zero-modes minus the number of fermionic zero-modes in a theory
containing SUSY. While SUSY constrains this difference to
be zero for any finite energy, at exactly zero energy the constraint is lifted.
Being a topological invariant, this index is highly stable, and is
in particular completely independent of temperature.

We note that traditionally in the study of topological phases,
topological indices result from properties of the bulk insulating phase, and dictate the edge
physics through the bulk-boundary correspondence. The topological invariant we study, on the other hand, is an explicit property
of the gapless edge theory itself, revealed by supersymmetry. We discuss the possibility of understanding it from the point of
view of the bulk topological quantum field theory, and highlight its connection to the entanglement spectrum
of a bipartition of the state in the bulk. Intriguingly,  while topological invariants are generically defined for
{\it non-interacting} symmetry protected topological phases, supersymmetry provides us with tools to define topological
invariants for {\it strongly interacting} RR states.

Clearly, any perturbation preserving supersymmetry does not alter the above structure. As we will see, SUSY is a robust property
of the $\nu=1/3$ Laughlin state. However, in other experimentally relevant states, given by filling factors of the form $3-\frac{k}{k+2}$, inter-mode
interaction terms generally break SUSY. However, imprints of SUSY may still be observed. In particular, for weak SUSY breaking
terms, the difference between the number of bosonic and fermionic states near zero energy is still given by the Witten
index. Furthermore, we discuss the possibility of SUSY re-emerging in the presence of edge reconstruction. Finally, we briefly discuss
the possibility of measuring the robust zero-modes in a small circular edge configuration.

{\it The system:} The system we study is made of a two-dimensional electron gas in the quantum
Hall regime. We will be interested in studying fermionic RR states at filling $\nu=\frac{k}{k+2}$. However, for
pedagogical reasons, we start from a different system made of two layers, the first contains bosons while the second contains
fermions. Both layers are in the quantum Hall regime, and we fix their densities such that the filling of the fermionic (bosonic)
layer is $\nu=1$ ($\nu=k/2$). While such a fermion-boson double layer system is beyond experimental reach, studying it will
provide a clear demonstration of the emergence of a supersymmetric low-energy sector on the edge. Furthermore, as we will later
argue, the edge theory of realistic fermionic RR states can be mapped to the supersymmetric theory on the edge of the
fermion-boson double layer. This will prove  the existence of SUSY on the edge of fermionic RR states.

Focusing first on the auxiliary fermion-boson double layer system, we explicitly write the edge theories of the two layers. The
fermionic layer is made of the trivial $\nu=1$ IQH state, whose edge contains a chiral free fermion field described by the
Hamiltonian $H_{\psi}=-iv\int dx\psi^\dagger \partial \psi$. The bosonic layer, on the other hand, is more complicated. It is
assumed to be in a bosonic RR state, whose edge realizes a strongly interacting $SU(2)_k$ CFT with central charge
$c=\frac{3k}{k+2}$.

The $SU(2)_k$ theory can be further decomposed into two mutually commuting sectors: a $U(1)$ charge mode with central charge $c_{U(1)}=1$
and a $SU(2)_{k}/U(1)$ theory with $c_{\mathbb{Z}_k}=2\frac{k-1}{k+2}$, describing $\mathbb{Z}_{k}$-parafermions.

The Hamiltonian describing the charge sector of the bosonic RR state is given by
\begin{equation}\label{eq:charge mode bosons}
H_{\rho}^b=\frac{v}{4\pi}\int dx\left(\partial_{x}\varphi_{\rho}^b\right)^{2},
\end{equation}
where $\varphi_\rho^b$ is a boson field.

The neutral parafermionic sector is an inherently strongly interacting CFT.
As the low-energy physics is independent of the microscopic representation,
 we can introduce a representation of the parafermionic CFT in terms of $k-1$ coupled bosons $\varphi_\sigma^\alpha$
 (with $\alpha=1,\cdots,k-1$). The $k$ different bosons (including the charge mode) satisfy the commutation relations
 \begin{equation} \label{eq:commutations }
[\varphi_\mu (x),\varphi_\nu (x')]=i\pi\delta^{\mu \nu}\text{sign}(x-x').
\end{equation}

 In terms of the neutral bosons, we write the Hamiltonian
\begin{align}\label{eq:Hamiltonian of parafermions}
 \!\! H_{\mathbb{Z}_k}=
  \frac{v}{2\pi(k+2)}\int dx\left[\left(\partial_{x}\vec{\varphi}_{\sigma}\right)^{2}+\sum_{a\neq b}e^{i\sqrt{2}(\vec{d}_{a}-\vec{d}_b)\cdot\vec{\varphi}_{\sigma}}\right],
\end{align}
where the $k$ vectors $\vec{d}_a$ (each of dimension $k-1$) satisfy the relations
\begin{equation}\label{eq:d}
\sum_{a=1}^k\vec{d}_{a}=0,\quad
\sum_{a=1}^k d_{a}^{\alpha}d_{a}^{\beta}= \delta_{\alpha\beta}, \quad
\vec{d}_{a}\cdot\vec{d}_{b}= \delta_{ab}-\frac{1}{k}.
\end{equation}

Indeed, the Hamiltonian $H_{\mathbb{Z}_k}$ describes a critical system, whose the low-energy sector coincides with the parafermionic sector of our edge theory \cite{Teo2014}.
We emphasize that the above bosonic representation is not unique, but simply
a useful one %
\footnote{We note, however, that one can devise a microscopic model for the Read-Rezayi state in terms of coupled wires, in which
case the bosons used here correspond to the bosonized degrees of freedom of physical wires.}. In terms of the above, we can write the parafermion operator as $\Psi=\sum_{a}e^{\sqrt{2}i\vec{d}_{a}\cdot\vec{\varphi}_{\sigma}}$.

 We can decompose the Hilbert space of the full fermion-boson double layer into the following two mutually commuting sectors: The first sector describes the total charge degrees of freedom, and is given by the Hamiltonian
\begin{align}\label{decomposition_susy1}
 H_{\rm U(1)}=&\frac{k}{k+2}H_{\rho}^b+\frac{2}{k+2}H_{\psi}-H_{\rm int},
\end{align}
with
\begin{equation}
 H_{\rm int}=\frac{v}{2(k+2)\pi}\sqrt{\frac{k}{2}}\int dx \psi^\dagger\psi \partial_x\varphi_\rho^b.
\end{equation}
The remaining sector is governed by the Hamiltonian
\begin{align}\label{decomposition_susy2}
 H_{SUSY}=&H_{\mathbb{Z}_k}+\frac{k}{k+2}H_{\psi}+\frac{2}{k+2}H_{\rho}^b+H_{\rm int}.
\end{align}
Note that $H_\psi+H_{\rho}^b+H_{\mathbb{Z}_k}=H_{\rm U(1)}+H_{SUSY}$.

As we demonstrate in the supplemental material using the properties of the $SU(2)_k$ and fermionic CFTs, the Hamiltonian
$H_{SUSY}$ exhibits an $\mathcal{N}=2$ SUSY. We show this by directly identifying two fermionic currents $G_1,G_2$ satisfying
the superconformal algebra.

In the above analysis we focused on the auxiliary fermion-boson system. Recall, however, that we are interested in a fermionic
RR state at filling $\nu=\frac{k}{k+2}$ (without an additional unrealistic bosonic subsystem). The Hamiltonian
describing such a fermionic RR a system is similar to the one describing the bosonic RR state. In particular,
the neutral sector, described by Eq. \ref{eq:Hamiltonian of parafermions}, remains unchanged. However, the charged degrees of
freedom are now described by the Hamiltonian
\begin{equation}\label{eq:charge mode fermions}
H_{\rho}^f=\frac{v}{4\pi}\frac{k+2}{k}\int dx\left(\partial_{x}\varphi_{\rho}\right)^{2},
\end{equation}
where $\varphi_\rho$ is a boson field satisfying
\begin{equation} \label{eq:commutations of charge modes fermion}
[\varphi_\rho (x),\varphi_\rho (x')]=i\pi\frac{k+2}{k}\text{sign}(x-x').
\end{equation}

In terms of these, the electron operator is given by $\psi_{el}^\dagger= \Psi e^{i\frac{k}{k+2}\varphi_\rho}$. This operator has
a scaling dimension of $3/2$, and a unit of electric charge. It can therefore be thought of as a dressed version of the
microscopic electron field, which commutes with the non-trivial bulk Hamiltonian.

Remarkably, the full Hamiltonian describing the fermionic RR state (given by
a combination of Eqs. (\ref{eq:Hamiltonian of parafermions}) and (\ref{eq:charge mode fermions})), can be mapped to the supersymmetric
Hamiltonian $H_{SUSY}$ studied in the fermion-boson auxiliary system.
The mapping between the neutral sector of the fermion-boson double layer and the fermionic RR state is shown explicitly in the supplemental material. The above line of arguments shows that the well established
fermionic RR state possesses $\mathcal{N}=2$ supersymmetry. Intriguingly, the fermionic current generating the SUSY transformations is given by the electron operator defined above (up to a
multiplicative constant).

{\it SUSY and its consequences:}
The presence of SUSY can be demonstrated explicitly by writing
two Hermitian
fermionic conserved currents, $G_1$ and $G_2$, satisfying the superconformal algebra shown in the supplemental material. In
the fermionic RR edge CFT, these two currents take the form \begin{align}
G_{1} & =\frac{1}{\sqrt{k+2}}\left(\psi_{el}^\dagger+\psi_{el}\right),\nonumber \\
G_{2} & =\frac{i}{\sqrt{k+2}}\left(\psi_{el}^{\dagger}-\psi_{el}\right).\label{eq:fermionic currents}
\end{align}
Note, in particular, that the above results apply to the $\nu=1/3$ Laughlin state (see supplemental material for the details of this simple case).

 By integrating over space, we get two fermionic charges, $Q_1,Q_2$, satisfying \footnote{We omit the contribution of the Casimir energy, related to the central charge
$c$, which vanishes in the long edge limit}
\begin{equation}\label{H_from_Q}
 Q_1^2=Q_2^2=H_{SUSY},
\end{equation}
with $\{Q_1,Q_2\}=0$, and  $H_{SUSY}=H_\rho^f +H_{\mathbb{Z}_k}$,  as shown explicitly in the supplemental material.

We further define a complex fermionic charge according to $Q=\frac{Q_1+iQ_2}{\sqrt{2}}=\sqrt{\frac{2v}{k+2}}\int dx \psi_{el}$. By definition, the complex charge satisfies
$Q^2=(Q^\dagger)^2=0$, and can be used to write the Hamiltonian in the convenient form
\begin{equation}\label{H_from_Q_complex}
H_{SUSY} = \frac{1}{2}\{Q^\dagger,Q\}.
\end{equation}

This simple structure allows us to study
the ground state of $H_{SUSY}$. Here we follow the arguments presented in
 Ref. \cite{Witten1982} to define an appropriate topological invariant.
From Eq. (\ref{H_from_Q_complex}) we see that $[H_{SUSY},Q]=0$. This implies that for each bosonic state $|\xi\rangle$ with energy $E>0$
\footnote{We note that any supersymmetric Hamiltonian is positive definite}, there is a fermionic state with the same energy, given by $|f\rangle\sim Q|\xi\rangle$.
To be more precise, assuming a normalized state $|\xi\rangle$, the normalized fermionic partner of this state is given explicitly by
$|f\rangle=\frac{1}{\sqrt{E}}Q|\xi\rangle$.

For the zero energy states, on the other hand, the relation between fermionic and
bosonic states is broken, and one generally has a different number of fermionic and bosonic zero-modes. To characterize this difference, it is useful to introduce the so called Witten index $W={\rm tr}(-1)^F$,
where $F=0$ for a bosonic state and $F=1$ for a fermionic state. We note that a natural realization of $W$ is given in terms of the angular
momentum operator $L^z$, such that $(-1)^F=\exp(2\pi i L^z)$. It is easy to see that the $W$ operator measures the difference
between the number of bosonic and fermionic states at zero energy, i.e.  $W=(N_{\rm B}-N_{\rm F})_{E=0}$. For our system
this difference has been calculated in Ref. \cite{Lerche1989} and is given by
\begin{align}
 W&=(N_{\rm B}-N_{\rm F})_{E=0}
 =\left\{
\begin{array}{ll}
      0 & k\quad\mbox{odd} \\
      1 & k\quad\mbox{even}
\end{array}\right..
\end{align}
As we argue below, this quantity is a topological invariant characterizing the RR edge theory \cite{Witten1982,Lerche1989}.
\begin{figure}[th]
\includegraphics[scale=0.35]{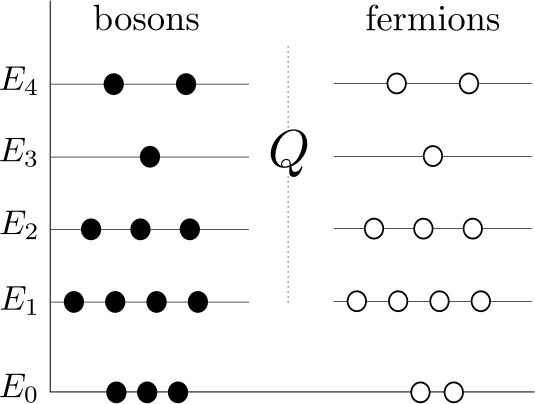}

\protect\caption{\label{fig:spectrum_susy} A generic spectrum of a supersymmetric theory. States at positive energies
come in degenerate Fermi-Bose pairs, related by the action of the SUSY charge $Q$. At zero energy, no such constraint exists and the number of fermionic and bosonic zero-modes can be different. The difference between the number of bosonic and fermionic states is called the Witten index $W$, and constitutes a topological invariant.}
\end{figure}

In order for the number of zero modes to change, a state must change its energy from zero to some positive value, or vice versa. However, since positive energy states come in Bose-Fermi pairs, the change in the number of bosonic and fermionic zero modes must be identical, meaning $W$ must remain fixed as long as SUSY is preserved. This prompts us to regard $W$ as a topological invariant.

The relation (\ref{H_from_Q_complex}) can be used to show that the zero modes of $H_{SUSY}$ are given by the states
that satisfy $Q|\xi\rangle=0$, but cannot be written as $|\xi\rangle= Q|\chi\rangle$ for some state $|\chi\rangle\neq0$ \cite{Witten1982}.
Clearly, since $Q^2=0$, any state of the form
$|\rho\rangle\equiv Q|\chi\rangle$ is annihilated by $Q$. Positive energy eigenstates $|\xi\rangle$ of $H_{SUSY}$ which are
annihilated by $Q$ can indeed be written as $|\xi\rangle=Q |\chi\rangle$, with $|\chi\rangle=\frac{1}{2E}Q^\dagger|\xi\rangle$.

The zero modes $|\rho\rangle$ of $H_{SUSY}$, on the other hand, are annihilated by $Q$ (this follows from
$\langle \rho|H_{SUSY}|\rho\rangle=\frac{1}{2}||Q|\rho\rangle||^2=0$), but cannot be written as $Q$ times some other state. This is so
because if $|\rho\rangle=Q|\mu\rangle$, then the states $|\rho\rangle$ and $|\mu\rangle$ have the same energy $E=0$ (as $[H_{SUSY},Q]=0$).
However, any state of zero energy satisfies $Q|\mu\rangle=0$, leading to a contradiction.

We refer to the space of solutions of $Q|\xi\rangle=0$ as the Kernel of the operator $Q$, or
$\text{Ker}(Q)$. Additionally, the space of states which can be written as $|\xi\rangle=Q |\chi\rangle$ for some state
$|\chi\rangle$ is referred to as the image of the operator Q, or $\text{Im}(Q)$. Using these definitions, the number of zero
modes is given by the dimension of the space spanned by states belonging to $\text{Ker}(Q)$ but not to $\text{Im}(Q)$, also
referred to as the cohomology of the operator $Q$ \cite{Witten1982}:

\begin{equation}
 N=(N_B+N_F)_{E=0}={\rm dim}\left(\frac{{\rm Ker} Q}{{\rm Im} Q}\right).
\end{equation}

In our case, $H_{SUSY}$ can be obtained by a coset decomposition of the Super-AKM algebra \cite{goddard1986,Kazama1989}
of $SU(2)_{k}$, into its $U(1)$ subalgebra, and the $\mathcal{N}=2$ SUSY sector $SU(2)_{k}/U(1)$. The number of zero modes
in this case is given by
\begin{equation}
 N=k+1,
\end{equation}
and is also known as the dimension of the chiral ring of the SUSY system \cite{Lerche1989}.

\begin{figure}[!ht]
\includegraphics[scale=0.3]{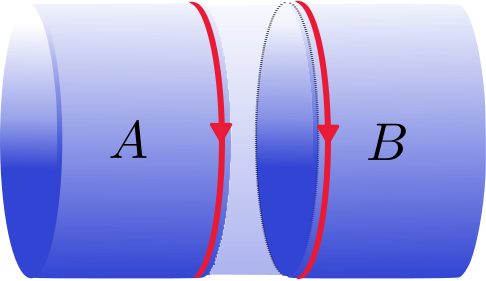}
\protect\caption{\label{fig:Entanglement_cut} Performing a cut in real space in the bulk of the quantum Hall state, here
displayed on the surface of a cylinder, reveals the entanglement spectrum, given by the thermal density matrix
of the chiral edge theory. This indicates that edge invariants $W$ and $N$ can be connected to characteristics of the bulk.}
\end{figure}

The above topological invariant  $W$, and the number of zero modes $N$ are associated with the edge of the Quantum Hall system,
but can be connected with properties of the bulk through the bulk-boundary correspondence. In particular, if we consider
a bipartition in the quantum Hall system (Fig \ref{fig:Entanglement_cut}), we will find that the entanglement Hamiltonian along
the boundary of the bipartition is precisely given by the Hamiltonian of the low-energy chiral edge theory \cite{Qi2012}, which
exhibits SUSY, and in particular displays $W$ and $N$ as characteristics of the spectrum. This prompts us to expect that
supersymmetry may arise directly from bulk properties.

{\it Physical perturbations:}
Being a symmetry which does not occur commonly in condensed matter systems, it is natural to ask
to which extent SUSY is robust. For $\nu=1/3$, any perturbation within the low-energy chiral Luttinger liquid theory merely
renormalizes the Fermi velocity. In this case, SUSY is indeed protected as long as coupling to other low-energy degrees of
freedom can be neglected.

The particle-hole conjugate of the RR states in the excited Landau Level, given by $\nu=3-\frac{k}{k+2}$ are prominent
candidates for describing the plateaus observed at $\nu=5/2$ and $12/5$. In this cases, the edge is described by three
co-propagating fermionic modes, and one counter propagating RR edge mode. Density-density interactions between the RR
edge mode and the fermionic channels generally break SUSY. However, we note that the zero modes associated with the
parafermionic sector remain unchanged, as the latter is charge neutral. Furthermore, if the SUSY breaking perturbations are
weak, the overall shape of the spectrum should weakly deviate from the supersymmetric spectrum presented in Fig.
\ref{fig:spectrum_susy}.

It was further shown in Ref. \cite{Bishara2008} that in the presence of disorder, such interactions induce an emergent $SU(2)_k$
algebra in the counter propagating sector. If edge reconstruction occurs (see, for example, Ref. \cite{Zhang2014}), an additional
counter-propagating fermionic mode emerges. As the analysis of the auxiliary fermion-boson double-layer system suggests, the emergent
$SU(2)_k$ theory, together with the fermionic channel, can again give rise to $\mathcal{N}=2$ SUSY for an appropriate choice of
parameters.

{\it Discussion:} In this manuscript, we have shown that quantum Hall states at filling $\nu=\frac{k}{k+2}$ constitutes a condensed matter realization of
an $\mathcal{N}=2$ supersymmetric conformal field theory. This allowed to use the Witten topological index, defined specifically for
supersymmetric theories, to count the difference between the bosonic and fermionic zero modes. We further discussed the
stability of the above against perturbations expected to occur in a physical realization. Remarkably, SUSY was found to be a particularly robust property of the Laughlin state at filling $\nu=1/3$.

General arguments dictate that the bipartite entanglement properties of the bulk of the system should also display the SUSY. Given this connection,
we expect that SUSY invariants could be discovered directly from the bulk low-energy topological field theory. The results presented here, and in particular the presence of SUSY on the edge, provide a strong indication that a
supersymmetric bulk theory, such as the one presented in Ref. \onlinecite{Tong2015}, are indeed adequate to study the the low-energy
physics of such quantum Hall states.

One can in principle measure the predicted zero-modes by creating a small circular edge, in which case the positive energy states
acquire an energy $E\propto1/L$, where $L$ is the circumference of the edge. Thus, the zero-modes become effectively isolated from the rest
of the spectrum. If, in addition, coupling to the external edge is taken into account, the zero modes are generally expected to
slightly deviate from zero energy, and can therefore be distinguished. By measuring the energy spectrum in this case, one can in principle
observe the $k+1$ zero modes.

It is interesting to consider if any of the ideas discussed here can be generalized to other Abelian states.
Given that the electron field in a $\nu=1/m$ theory has conformal dimension $m/2$, the trivial generalization of taking those
fields to be the generators of supersymmetry clearly does not work. On the other hand, the mathematical structure of such abelian states
is similar, making it natural to expect the ideas presented here extend to those states as well.

\smallskip
\begin{acknowledgments}
We would like to thank M. Barkeshli, Y. Fuji, P. Liendo, J.Park, Y. Oreg and Y.Gefen for informative discussions. This work was supported by the
Feinberg School at WIS, the Israel Science Foundation (ISF),
the European Research Council under the European Community's Seventh
Framework Program (FP7/2007-2013)/ERC Grant agreement No. 340210, DFG CRC TR 183, the Binational Science
Foundation (BSF), and the Adams Fellowship Program of the Israel Academy of Sciences and Humanities.\end{acknowledgments}

\bibliographystyle{apsrev4-1}
%


\onecolumngrid
\section*{\large{Supplemental Material}}

\section{From the Hamiltonian to the energy-momentum tensor}

In this section, we demonstrate how the Hamiltonian can be obtained from the
corresponding energy-momentum tensor. As we will use the CFT formulation to demonstrate the existence of $\mathcal{N}=2$ SUSY, the energy-momentum tensor will play a central role in our discussion.

If we impose periodic boundary conditions in the $x$- direction (i.e.,
we take a circular edge), the coordinates $\tau$ and $x$ define
a cylinder.
As is well known, one can then apply a conformal transformation that
goes from the cylinder to a plane:
\[
z=e^{2\pi\left(v\tau+ix\right)/L},
\]
where $L$ is the circumference of the edge. Time ordering on the
physical cylinder corresponds to radial ordering on the transformed
plane. While the resulting plane has no immediate physical significance,
working on it greatly simplifies the analysis of conformal field theories.

Defining the stress-energy tensor $T(z)$ on the plane, one can extract
the Hamiltonian according to
\[
H=\frac{2\pi v}{L}L_{0}\quad \mbox{with}\quad L_0=\frac{1}{2\pi i}\int_{C}dzzT(z),
\]
where $C$ is a circular contour going around the origin. The connection
to the physical cylinder is seen by performing the transformation
from $z$ to $x$ (working in fixed time $\tau$), and recalling that
conformal fields transform according to
\[
\phi(z)\rightarrow\phi(x)\left(\frac{\partial x}{\partial z}\right)^{h}.
\]

For example, using the above, we will find that the supersymmetric Hamiltonian $H_{SUSY}$ (\ref{decomposition_susy2}) obtained from the
auxiliary fermion-boson system, corresponds to the zero mode of the energy-momentum tensor (see also Sec. \ref{app:N2SUSY})
\begin{equation}\label{emt_susy}
 T_{SUSY}(z)=\frac{1}{k+2}\left[-\frac{k}{2}(:\psi^\dagger\partial\psi:-:\partial\psi^\dagger\psi:)-2\mathcal{J}^3:\psi^\dagger\psi:+\frac{1}{2}(:\mathcal{J}^+\mathcal{J}^-:+:\mathcal{J}^-\mathcal{J}^+:)\right](z).
\end{equation}
where $\mathcal{J}^3,\mathcal{J}^\pm$ are the ${\rm SU(2)}_k$ currents defined in the next section.
As we will show in the next sections, this theory exhibits satisfied an $\mathcal{N}=2$ superconformal algebra.

\section{Operator Product Expansion (OPE)}
In this section we list the OPEs necessary to derive the $\mathcal{N}=2$ SUSY algebra. Using the OPEs of a free complex fermion
and boson fields as building blocks, other relations can be obtained. In particular, we explicitly show how the OPEs of the ${\rm SU}(2)$
Affine Kac-Moody (AKM) algebra can be represented in terms of bosonic vertex operators.

\subsection{Complex Fermion}
The OPE of a complex fermion is given by
\begin{equation}\label{OPE_psid_psi}
 \psi^\dagger(z)\psi(w)\sim\frac{1}{z-w}+:\psi^\dagger\psi(w):+(z-w):\partial\psi^\dagger\psi(w):+\dots,
\end{equation}
where the ellipsis represent higher powers of $(z-w)$. The normal ordering of the field $\mathcal{O}$ is represented as usual
by the $:\mathcal{O}:$, and correspond to the substraction of all the singular terms in the limit where the arguments
coincide.
For future reference, we also include here the OPE of Majorana fields, obtained from the complex
fermion field by $\psi_1=(\psi^\dagger+\psi)/\sqrt{2}$ and $\psi_2=(\psi^\dagger-\psi)/\sqrt{2}i$. The OPE of a pair of Majorana fields takes the form
\begin{equation}\label{MajoranaOPE}
 \psi_a(z)\psi_b(w)\sim \frac{\delta_{ab}}{z-w}+:\psi_a\psi_b(w):+(z-w):\partial\psi_a\psi_b(w):+\dots
\end{equation}

 An additional useful relation is
\begin{equation}
 -:\psi_a\partial\psi_a(z):\psi_a(w)\sim \frac{\psi_a(w)}{(z-w)^2}+\frac{2\partial\psi_a(w)}{z-w}+\dots.
\end{equation}

\subsection{Free Boson}
The OPE between bosonic fields $\phi_a$ is given by $\partial\phi_a(z)\partial\phi_b(w)\sim -\delta_{ab}/(z-w)^2+\dots$. This corresponds to the familiar commutation relation of bosonic fields in the Luttinger liquid theory, expressed in the language of CFT
\cite{Ketov1995,*DiFrancesco1997,*Fradkin2013}.

Using this OPE and the Baker-Hausdorff-Campbell relation, the OPE between two vertex operators reads
\begin{equation}
 e^{i\alpha\phi_a(z)}e^{i\beta\phi_b(\omega)}=\frac{1}{(z-w)^{-\alpha\beta}}\exp\left(i(\alpha\phi_a(z)+\beta\phi_a(w))\right)\delta_{ab}.
\end{equation}
Using the Wick theorem and the OPE between bosonic fields, it is also straightforward to show that
\begin{equation}
 \partial\phi_a(z)e^{i\alpha\phi_b(z)}\sim -i\alpha\frac{e^{i\alpha\phi_a(w)}}{z-w}\delta_{ab}+\dots\quad\mbox{and}\quad
 -(\partial\phi_a(z))^2e^{i\alpha\phi_b(z)}\sim \left(\alpha^2\frac{e^{i\alpha\phi_a(w)}}{(z-w)^2}
 +\frac{\partial(e^{i\alpha\phi_a(w)})}{z-w}\right)\delta_{ab}+\dots
\end{equation}

\subsection{${\rm SU}(2)_k$ currents}
The previous relations allow us to construct a representation of the ${\rm SU}(2)$ Affine Kac-Moody (AKM) algebra at level one.
Defining $J^{\pm}_a=e^{i\pm\sqrt{2}\phi_a}$ and $J^3_a=\frac{i}{\sqrt{2}}\partial\phi_a$ and using the relations between
vertex operators of bosonic fields discussed above, we find
\begin{equation}\label{AKMlevel1}
 J^3_a(z)J^3_b(w)\sim\frac{1/2\delta_{ab}}{(z-w)^2}+\dots,\quad J^+_a(z)J^{-}_b(w)\sim\frac{\delta_{ab}}{(z-w)^2}+
 \frac{2J_a^3(w)\delta_{ab}}{z-w}+\dots,\quad J^3_a(z)J^{\pm}_b(w)\sim \frac{\pm\delta_{ab}J_a^{\pm}(w)}{z-w}+\dots
\end{equation}
Defining $J^1=(J^++J^-)/2$ and $J^2=(J^+-J^-)/2i$, the previous relations can be brought into the compact form
\begin{equation}
 J_a^{m}(z)J_b^{n}(w)\sim\left(\frac{1/2}{(z-w)^2}\delta^{mn}+\frac{i\varepsilon^{mnl}J_a^l(w)}{z-w}\right)\delta_{ab}+\dots,
\end{equation}
where the sum over repeated indices is assumed. The ${\rm SU}(2)$ indices $(m,n,l)$ run from 1 to 3 and $\varepsilon^{mnl}$
is the Levi-Civita antisymmetric tensor, which parametrizes the structure constants of ${\rm SU}(2)$.

Adding $k$ different mutually commuting currents, we obtain a representation of the ${\rm SU}(2)$ AKM algebra at level $k$.
To be specific, defining $\mathcal{J}^m=\sum_{a=1}^kJ_a^m$, we find
\begin{equation}\label{AKMOPE}
  \mathcal{J}^{m}(z)\mathcal{J}^{n}(w)\sim\frac{k/2}{(z-w)^2}\delta^{mn}+\frac{i\varepsilon^{mnl}\mathcal{J}^l(w)}{z-w}
  +:\mathcal{J}^m\mathcal{J}^n(w):+\frac{i}{2}\varepsilon^{mnl}\partial\mathcal{J}^l(w)+\dots
\end{equation}
where we have included the first non singular term. In this representation of the ${\rm SU}(2)_k$  currents, the normal ordered product
appearing above is $:\mathcal{J}^+\mathcal{J}^-(w):\equiv\sum_{a\neq b} J_a^+J^-_b-\sum_a(\partial\phi_a)^2$, and similarly for
the other combinations.

For future use, it is useful to work in a basis of charge and neutral degrees of freedom. This can be done by defining
$\phi_a=\frac{1}{\sqrt{k}}\varphi_\rho+\vec{d}_a\cdot\vec{\varphi}_\sigma$, with the $\vec{d}$-vectors defined in Eq. \ref{eq:d} of the main text.
In terms of these, the currents take the form
\begin{equation}
 \mathcal{J}^{\pm}=\sum_{a=1}^ke^{\pm i\left(\sqrt{\frac{2}{k}}\varphi_\rho+\sqrt{2}\vec{d}_a\cdot\vec{\varphi}_\sigma\right)}\quad
 \mbox{and}\quad
 \mathcal{J}^3=i\sqrt{\frac{k}{2}}\partial\varphi_\rho.
\end{equation}

 When computing OPEs of product of fields, it is generally important to keep terms
that, although are nonsingular within a single OPE, multiplied by singular terms coming from other fields could still give a
nontrivial (i.e. singular) contribution. An example of this occurs in the computation of the OPE
\begin{eqnarray}\nonumber
 -2:\psi^\dagger\psi:\mathcal{J}^3(z)[\psi\mathcal{J}^++\psi^\dagger\mathcal{J}^-](w)&\sim
 \frac{2}{z-w}\left[\psi(z)\left(\frac{\mathcal{J}^+(w)}{z-w}+
  \frac{1}{2}\partial\mathcal{J}^+(w)\right)+\psi^\dagger(z)\left(\frac{\mathcal{J}^-(w)}{z-w}+
  \frac{1}{2}\partial\mathcal{J}^-(w)\right)\right]+\dots\\
&\sim  2\left[\frac{[\psi\mathcal{J}^++\psi^\dagger\mathcal{J}^-](w)}{(z-w)^2}+\frac{[(\partial\psi)\mathcal{J}^++(\partial\psi^\dagger)\mathcal{J}^-](w)}{z-w}+\frac{1}{2}\frac{[\psi\partial\mathcal{J}^++\psi^\dagger\partial\mathcal{J}^-](w)}{z-w}\right]+\dots
\end{eqnarray}
 here the top equation is obtained by using (\ref{OPE_psid_psi}) and (\ref{AKMOPE}). The bottom result is obtained expanding
the fields at $z$ in Taylor series, i.e $\psi(z)=\psi(w)+(z-w)\partial\psi(w)+\dots$.

\section{$\mathcal{N}=2$ Superconformal Algebra}\label{app:N2SUSY}

As we discussed in the main text, to explicitly show the supersymmetric structure of the RR theory, we first study an auxiliary fermion-boson double layer, whose edge consists
of a chiral fermion coupled to an ${\rm SU}(2)_k$ CFT (which corresponds to a bosonic Read-Rezayi state). Using the OPEs
introduced previously, we show that the neutral sector of this fermionic + bosonic theory satisfies an $\mathcal{N}=2$ superconformal algebra.
Finally, by bosonizing the fermion field, we show that the neutral part can
be mapped to the fermionic RR edge CFT.

First, we identify the $U(1)$ current associated with the total charge in the boson-fermion double-layer:
\begin{equation}\label{app:commuting}
 J_{U(1)}(z)=\frac{1}{\sqrt{k+2}}(\mathcal{J}(z)+:\psi^\dagger\psi:(z)).
\end{equation}
The energy-momentum tensor associated with the total charge degrees of freedom, $T_{U(1)}$, is given by the OPE of $J_{U(1)}$ with itself.
Clearly, the remaining part, $T_{SUSY}$, of the total energy-momentum tensor describes neutral degrees of freedom. It is defined such that
\begin{equation}
 T_{fb}=T_\psi+T_{\rho}^b+T_{\mathbb{Z}_k}=T_{U(1)}+T_{SUSY}
\end{equation}
We therefore find that the energy-momentum tensor corresponding to the neutral sector of the auxiliary theory is given by
\begin{equation}\label{emt_susy}
 T_{SUSY}(z)=\frac{1}{k+2}\left[-\frac{k}{2}(:\psi^\dagger\partial\psi:-:\partial\psi^\dagger\psi:)-2\mathcal{J}^3:\psi^\dagger\psi:+\frac{1}{2}(:\mathcal{J}^+\mathcal{J}^-:+:\mathcal{J}^-\mathcal{J}^+:)\right](z).
\end{equation}
We will find below that the OPE between $T_{U(1)}$ and $T_{SUSY}$ is non-singular.

Using the OPEs outlined in the previous section, it is possible to show that the energy-momentum tensor $T_{SUSY}$ satisfies the full
$\mathcal{N}=2$ superconformal algebra \cite{Kazama1989}, which reads
\begin{eqnarray}\label{SUSY_algebra}
 T(z)T(w)&\sim&\frac{c/2}{(z-w)^4}+\frac{2T(w)}{(z-w)^2}+\frac{\partial T(w)}{z-w},\\
 T(z)G^\alpha(w)&\sim&\frac{\frac{3}{2}G^\alpha(w)}{(z-w)^2}+\frac{\partial G^\alpha(w)}{z-w},\\\label{GG}
 G^\alpha(z)G^\beta(w)&\sim&\left[\frac{\frac{2}{3}c}{(z-w)^3}+\frac{2T(w)}{z-w}\right]\delta^{\alpha\beta}+i\left[\frac{2J(w)}{(z-w)^2}+\frac{\partial J(w)}{z-w}\right]\epsilon^{\alpha\beta},\\
 T(z)J(w)\sim \frac{J(w)}{(z-w)^2}+\frac{\partial J(w)}{z-w},&\quad&
 J(z)G^\alpha(w)\sim i\epsilon^{\alpha\beta} \frac{G^{\beta}(w)}{z-w},\quad J(z)J(w)\sim\frac{c/3}{(z-w)^2}.\label{SUSY_JG}
\end{eqnarray}
where $G^1(z)=\frac{1}{\sqrt{k+2}}(\psi(z)\mathcal{J}^+(z)+\psi^\dagger(z)\mathcal{J}^-(z)),G^2(z)=\frac{i}{\sqrt{k+2}}(\psi^\dagger(z)\mathcal{J}^-(z)-\psi(z)\mathcal{J}^+(z))$
are the two fermionic currents and $J(z)=\frac{2}{k+2}[\mathcal{J}^3(z)-\frac{k}{2}\psi^\dagger\psi]$ is a $U(1)$ current. The central charge here is $c=3k/(k+2)$. The tensor $\epsilon^{\alpha\beta}$ $(\alpha,\beta=1,2$) is antisymmetric with $\epsilon^{12}=1$.

To illustrate the results above, we compute explicitly an example of the OPE between two fermionic currents
\begin{eqnarray}\nonumber
 &G^1(z)G^1(w)\sim\frac{1}{k+2}[\psi(z)\psi^\dagger(w)\mathcal{J}^+(z)\mathcal{J}^-(w)+\psi^\dagger(z)\psi(w)\mathcal{J}^-(z)\mathcal{J}^+(w)]&\\\nonumber
 &\sim\frac{1}{k+2}\left(\frac{1}{z-w}+:\psi\psi^\dagger(w):+(z-w):\partial\psi\psi^\dagger(w):\right)\left(\frac{k}{(z-w)^2}+\frac{2\mathcal{J}^3(w)}{z-w}+:\mathcal{J}^+\mathcal{J}^-(w):+i\partial\mathcal{J}^3(w)\right)&\\\nonumber
 &+\frac{1}{k+2}\left(\frac{1}{z-w}+:\psi^\dagger\psi(w):+(z-w):\partial\psi^\dagger\psi(w):\right)\left(\frac{k}{(z-w)^2}-\frac{2\mathcal{J}^3(w)}{z-w}+:\mathcal{J}^-\mathcal{J}^+(w):-i\partial\mathcal{J}^3(w)\right).&
\end{eqnarray}
Here we have used the OPE of $\psi^\dagger(z)\psi(w)$ given in (\ref{OPE_psid_psi}) and the one analogous for $\psi(z)\psi^\dagger(w)$.
The OPE between two ${\rm SU}(2)_k$ currents is given by (\ref{AKMOPE}). Rearranging and keeping the singular terms, we find
\begin{equation}\nonumber
 G^1(z)G^1(w)\sim\frac{\frac{2k}{k+2}}{(z-w)^3}+\frac{\frac{2}{k+2}}{z-w}\left[-\frac{k}{2}(:\psi^\dagger\partial\psi:-:\partial\psi^\dagger\psi:)-2\mathcal{J}^3:\psi^\dagger\psi:+\frac{1}{2}(:\mathcal{J}^+\mathcal{J}^-:+:\mathcal{J}^-\mathcal{J}^+:)\right](w),
\end{equation}
where we recognize the first term on the right hand side of (\ref{GG}) with $T=T_{SUSY}$  (\ref{emt_susy}). All the other relations (\ref{SUSY_algebra}-\ref{SUSY_JG})
follow in a similar way.

To make a connection with the fermionic RR state, it is illuminating to write explicitly the neutral and
charged sectors of the auxiliary theory. To do this, we use the bosonic vertex representation of the ${\rm SU(2)}_k$ currents
\begin{equation}
 \mathcal{J}^{\pm}=\sum_{a=1}^ke^{\pm i\left(\sqrt{\frac{2}{k}}\varphi_\rho^b+\sqrt{2}\vec{d}_a\cdot\vec{\varphi}_\sigma\right)}\quad
 \mbox{and}\quad
 \mathcal{J}^3=i\sqrt{\frac{k}{2}}\partial\varphi_\rho^b.
\end{equation}
where $\varphi_\rho^b$ is a charged field (generating a ${\rm U(1)}$ subalgebra) with OPE $\partial\varphi_\rho^b(z)\partial\varphi_\rho^b(w)\sim-1/(z-w)^2$, and $\vec{\varphi}_\sigma$ is a vector of
neutral fields. The components of the neutral
bosonic fields that define the parafermionic sector satisfy $\varphi_\sigma^\alpha(z)\varphi_\sigma^\beta(w)\sim\ln(z-w)$.
Replacing these currents in (\ref{emt_susy}), we have
\begin{equation}\label{emt_susy_2}
 T_{SUSY}(z)=\frac{1}{k+2}\left[-\frac{k}{2}(:\psi^\dagger\partial\psi:-:\partial\psi^\dagger\psi:)-2i\sqrt{\frac{k}{2}}\partial\varphi_\rho^b:\psi^\dagger\psi:-(\partial\varphi_\rho^b)^2-(\partial\vec{\varphi}_\sigma)^2+\sum_{a\neq b}e^{i\sqrt{2}(\vec{d}_a-\vec{d}_b)\cdot\vec{\varphi}_\sigma}\right].
\end{equation}
To finally show the connection of this energy-momentum tensor with the RR edge CFT, we bosonize the fermion operator
$\psi^\dagger=e^{i\phi_1}$ with OPE $\partial\phi_1(z)\partial\phi_1(w)\sim-1/(z-w)^2$, leading to
\begin{equation}\label{emt_susy_3}
 T_{SUSY}(z)=\frac{1}{k+2}\left[-\frac{k}{2}(\partial\phi_1)^2+2\sqrt{\frac{k}{2}}\partial\varphi_\rho^b\partial\phi_1-(\partial\varphi_\rho^b)^2-(\partial\vec{\varphi}_\sigma)^2+\sum_{a\neq b}e^{i\sqrt{2}(\vec{d}_a-\vec{d}_b)\cdot\vec{\varphi}_\sigma}\right].
\end{equation}
Defining
\begin{equation}
 \varphi_\rho=\frac{k}{k+2}\left(\phi_1-\sqrt{\frac{2}{k}}\varphi_\rho^b\right),
\end{equation}
the auxiliary energy-momentum tensor becomes
\begin{equation}\label{SET_RR}
 T_{SUSY}(z)=-\frac{k+2}{2k}(\partial\varphi_\rho)^2-\frac{1}{k+2}\left((\partial\vec{\varphi}_\sigma)^2-\sum_{a\neq b}e^{i\sqrt{2}(\vec{d}_a-\vec{d}_b)\cdot\vec{\varphi}_\sigma}\right)\equiv T_{\rm RR}(z),
\end{equation}
which indeed coincides with the energy-momentum tensor of the RR edge theory. It follows from the definition that $\partial\varphi_\rho(z)\partial\varphi_\rho(w)\sim-\frac{k}{k+2}1/(z-w)^2$ so we identify
this field with the charge mode of the RR state.

We point out that the central charge of the auxiliary fermion-boson double-layer, given by $c_{fb}=\frac{3k}{k+2}+1$, is larger
than the central charge of the fermionic RR state, $c_{RR}=\frac{3k}{k+2}$. Indeed, the auxiliary system consists of an additional
total charge degrees of freedom, given by
\begin{equation}
 \varphi_{tot}=\frac{1}{\sqrt{k+2}}\left(\phi_1+\sqrt{\frac{k}{2}}\varphi_\rho^b\right).
\end{equation}
However, this mode commutes with the energy-momentum tensor $T_{SUSY}$ defined above. We can write the $U(1)$ current $J_{U(1)}$ in terms of this field as $J_{U(1)}=i\partial\varphi_{tot}$.  Given that $J_{U(1)}$ and the fermionic currents $\mathcal{G}^{1,2}$
do not have singular terms in their OPE, and that the fermionic currents generate $T_{SUSY}$ (see Eq. \ref{GG}), it follows that $H_{U(1)}$ and
$H_{SUSY}$, being zero modes of their corresponding energy-momentum tensors, commute.

We therefore emphasize that the above mapping is between the fermionic RR system and the neutral sector of the fermion-boson double layer,
rather than the entire theory.

The complex supersymmetric current operator $\mathcal{G}^\pm=\frac{1}{\sqrt{2}}(G^1\pm iG^2)=\sqrt{\frac{2}{k+2}}\psi^{\mp}\mathcal{J}^\pm=\sqrt{\frac{2}{k+2}}e^{\mp i\frac{k+2}{k}\varphi_\rho}\Psi^\pm$ is
simply the physical (annihilation/creation) electron operator, which indeed has conformal dimension $3/2$. The fermionic charge, generating the SUSY transformations, is given by the space
integral of $\mathcal{G}$. The OPE between $\mathcal{G}^+$ and $\mathcal{G}^-$ is obtained from (\ref{GG})
\begin{equation}\label{GG_complex}
 \mathcal{G}^-(z)\mathcal{G}^+(w)\sim\frac{\frac{2}{3}c}{(z-w)^3}+\frac{2T(w)}{z-w}-\frac{2J(w)}{(z-w)^2}-\frac{\partial J(w)}{z-w}\quad\mbox{and}\quad \mathcal{G}^+(z)\mathcal{G}^+(w)\sim \mathcal{G}^-(z)\mathcal{G}^-(w)\sim 0.
\end{equation}

\subsection{An explicit analysis of the Laughlin state $k=1$, $\nu=1/3$}

The simplest and most prominent example of a FQH edge theory endowed with $\mathcal{N}=2$ SUSY is the $\nu=1/3$ Laughlin state.
This filling fraction is realized in the RR series by taking $k=1$. In this case, the parafermion sector vanishes and the theory becomes Abelian. To illustrate the general results described above, we now explicitly demonstrate how they arise this simple case. In particular, we show that the SUSY algebra appears naturally in this case by making
use of the vertex OPE.

The electron operator is given by $\psi_{\rm el}^\dagger=e^{3i\varphi_\rho}$ and the fermionic current operator
is given by $\mathcal{G}^+=\sqrt{\frac{2}{3}}\psi_{\rm el}$. The OPE of two supercurrent operators is then
\begin{equation}\nonumber
 \mathcal{G^-}(z)\mathcal{G^+}(w)=\frac{2}{3}\psi^\dagger_{\rm el}(z)\psi_{\rm el}(w)=\frac{2/3}{(z-w)^3}e^{3i(\varphi_\rho(z)-\varphi_\rho(w))},
\end{equation}
where in the last equality we have used that $\varphi_\rho(z)\varphi_\rho(w)\sim 1/3\ln(z-w)$. Expanding the difference in the
exponent in Taylor series around $w$, we have
\begin{eqnarray}\nonumber
 \mathcal{G^-}(z)\mathcal{G^+}(w)&=&\frac{2/3}{(z-w)^3}\exp\left(3i\sum_{n\geq 1}\frac{(z-w)^n}{n!}\partial^n\varphi_\rho\right)\\\nonumber
 &=&\frac{2/3}{(z-w)^3}\left(1+3i(z-w)\partial\varphi_\rho-\frac{9}{2}(z-w)^2(\partial\varphi_\rho)^2+\frac{3i}{2}(z-w)^2\partial^2\varphi_\rho+\dots\right)\\
 &=&\frac{2/3}{(z-w)^3}+\frac{2(i\partial\varphi_\rho)}{(z-w)^2}+\frac{2(-\frac{3}{2}(\partial\varphi_\rho)^2)}{(z-w)}+\frac{\partial(i\partial\varphi_\rho)}{(z-w)}+\dots
\end{eqnarray}
which corresponds to the expression (\ref{GG_complex}) for $c=1$, $T(z)=-\frac{3}{2}(\partial\varphi_\rho)^2$ and
$J(z)=-i\partial\varphi_\rho$. The different OPEs that give rise to the full superconformal algebra
(\ref{SUSY_algebra}-\ref{SUSY_JG}) can be obtained in a similar way.

\subsection{Mode expansion}
Using the previous OPEs, it is possible to find the (anti)-commutation relations between the different modes of the fields.
These modes are given in the Laurent expansion of the fields i.e.
\begin{eqnarray}
  T(z)=\sum_{n\in \mathbb{Z}}\frac{L_n}{z^{n+2}},\quad G(z)=\sum_{n\in \mathbb{Z}}\frac{G_n}{z^{n+\frac{3}{2}}}&\quad&\mbox{and}\quad
  J(z)=\sum_{n\in \mathbb{Z}}\frac{J_n}{z^{n+1}},\\
  L_m=\oint_0\frac{dz}{2\pi i}z^{m+1}T(z),\quad G_m=\oint_0\frac{dz}{2\pi i}z^{m+\frac{1}{2}}G(z)&\quad&\mbox{and}\quad
  J_m=\oint_0\frac{dz}{2\pi i}z^{m}J(z),
\end{eqnarray}
The (anti)-commutation relations are (in the Ramond sector)
\begin{eqnarray}
[L_n,L_m]&=&(n-m)L_{n+m}+\frac{c}{12}n(n^2-1)\delta_{n+m} \\
\left[L_n,G_m\right]&=& \left(\frac{n}{2}-m\right)G_{n+m}\\
 \{G^\alpha_n,G^\beta_m\}&=&\delta^{\alpha\beta}\left(\frac{c}{3}\left(n^2-\frac{1}{4}\right)\delta_{n+m}+2L_{n+m}\right)+i\epsilon^{\alpha\beta}(n-m)J_{n+m},\\
\left[L_n,J_m\right]=-mJ_{n+m},&\quad& [J_n,G_m^\alpha]=i\epsilon^{\alpha\beta}G^\beta_{n+m},\quad [J_n,J_m]=\frac{c}{3}n\delta_{m+n}.
\end{eqnarray}
The combinations $\mathcal{G}^\pm_n=\frac{1}{\sqrt{2}}(G^1_n\pm i G^2_n)$ correspond to the electron operator, which is also the
fermionic current, with anti-commutation relations
\begin{equation}\label{Hamiltonian_from_Q}
 \{\mathcal{G}^+_n,\mathcal{G}^-_m\}=\frac{c}{3}\left(n^2-\frac{1}{4}\right)\delta_{n+m}+2L_{n+m}+(n-m)J_{n+m},\quad\mbox{and}\quad\{\mathcal{G}^+_n,\mathcal{G}^+_m\}=\{\mathcal{G}^-_n,\mathcal{G}^-_m\}=0.
\end{equation}
The SUSY charge Q, defined in the main text, corresponds to $\sqrt{\frac{2\pi v}{L}}\mathcal{G}^+_0$. Plugging this in, we find that
\begin{equation}\label{Hamiltonian_from_Q2}
 \frac{1}{2}\{Q,Q^\dagger\}=H_{SUSY}.
\end{equation}

\twocolumngrid

\end{document}